\documentclass[]{fairmeta}

\usepackage{amsmath,amsfonts,bm}

\def\eqref#1{equation~\ref{#1}}

\def\1{\bm{1}}

\DeclareMathAlphabet{\mathsfit}{\encodingdefault}{\sfdefault}{m}{sl}
\SetMathAlphabet{\mathsfit}{bold}{\encodingdefault}{\sfdefault}{bx}{n}

\usepackage{amssymb}
\usepackage{enumitem}
\usepackage{tabularx}
\usepackage{adjustbox}
\usepackage{float}
\usepackage{array}
\usepackage[scaled=0.85]{sourcecodepro}
\usepackage{fancyvrb}

\makeatletter
\renewcommand\authorformat[2][]{{\sffamily \bfseries #2$^{\scriptscriptstyle #1}$}}
\makeatother

\title{\fontsize{15}{18}\selectfont NeuralSet: A High-Performing Python Package for Neuro-AI}

\author[1]{Jean-R\'emi King}

\author[1,2,*]{Corentin Bel}
\author[3,*]{Linnea Evanson}
\author[3,5,*]{Julien Gadonneix}
\author[1,4,*]{Sophia Houhamdi}
\author[1,4,*]{Jarod L\'evy}
\author[1,2,*]{Josephine Raugel}
\author[3,5,*]{Andrea Santos Revilla}
\author[2,3,*]{Mingfang (Lucy) Zhang}

\author[6]{Julie Bonnaire}
\author[1]{Charlotte Caucheteux}
\author[1]{Alexandre D\'efossez}
\author[1]{Th\'eo Desbordes}
\author[2]{Pablo Diego-Sim\'on}
\author[1]{Shubh Khanna}
\author[1]{Juliette Millet}
\author[1]{Pierre Orhan}
\author[1]{Saarang Panchavati}
\author[1]{Antoine Ratouchniak}
\author[1]{Alexis Thual}

\author[1,\dagger]{Teon L.\ Brooks}
\author[1,\dagger]{Katelyn Begany}
\author[1,\dagger]{Yohann Benchetrit}
\author[1,\dagger]{Marl\`ene Careil}
\author[1,\dagger]{Hubert Banville}
\author[1,\dagger]{St\'ephane d'Ascoli}
\author[1,\dagger]{Simon Dahan}

\author[1]{J\'er\'emy Rapin}

\affiliation[1]{FAIR, Meta}
\affiliation[2]{\'Ecole Normale Sup\'erieure - PSL Universit\'e, Paris}
\affiliation[3]{Foundation Adolphe de Rothschild Hospital, Paris}
\affiliation[4]{MIND, Inria}
\affiliation[5]{Université Paris Cité}
\affiliation[6]{Neurospin, CEA, Gif sur Yvette}

\contribution[*]{These authors contributed equally.}
\contribution[\dagger]{These senior authors contributed equally.}

\metadata[Code]{\url{https://github.com/facebookresearch/neuroai/}}
\correspondence{\email{jeanremi@meta.com}, \email{jrapin@meta.com}}

\abstract{
Artificial intelligence (AI) is increasingly central to understanding how the 
brain processes information. 
However, the integration of neuroscience and modern AI is bottlenecked by a 
fragmented software ecosystem. 
Current tools are siloed by recording modality and optimized for small-scale, 
in-memory workflows, limiting the use of massive, naturalistic datasets. 
Here, we introduce NeuralSet, a Python framework that efficiently unifies 
the processing of diverse neural recordings (including fMRI, M/EEG, and spikes) 
and complex experimental stimuli (such as text, audio, and video). 
By decoupling experimental metadata from lazy, memory-efficient data extraction, 
NeuralSet harmonizes standard neuroscientific preprocessing pipelines with pretrained deep learning embeddings. 
This approach provides a single PyTorch-ready interface 
that scales seamlessly from local prototyping to high-performance cluster execution. 
By eliminating manual data wrangling and ensuring full computational provenance, 
NeuralSet establishes a scalable, unified infrastructure for the next generation 
of neuro-AI research.
}

\begin{document}
\maketitle

\definecolor{cEEG}{HTML}{2980B9}
\definecolor{cMEG}{HTML}{16A085}
\definecolor{cFMRI}{HTML}{27AE60}
\definecolor{cText}{HTML}{D4A017}
\definecolor{cImage}{HTML}{E67E22}
\definecolor{cAudio}{HTML}{C0392B}
\definecolor{cPipe}{HTML}{3C3C3C}
\tcbset{extcard/.style={
  arc=1.5pt, boxrule=0pt,
  left=2pt, right=2pt,
  top=0pt, bottom=0pt,
  before skip=0pt, after skip=0pt,
  middle=0pt,
}}
\newcommand{\codeline}[2]{\nointerlineskip\noindent\colorbox{#1}{\makebox[\dimexpr\textwidth-4pt][l]{\footnotesize\ttfamily\strut #2}}\par}

\begin{figure}[H]
\vspace{-24pt}
\centering
\includegraphics[width=\textwidth]{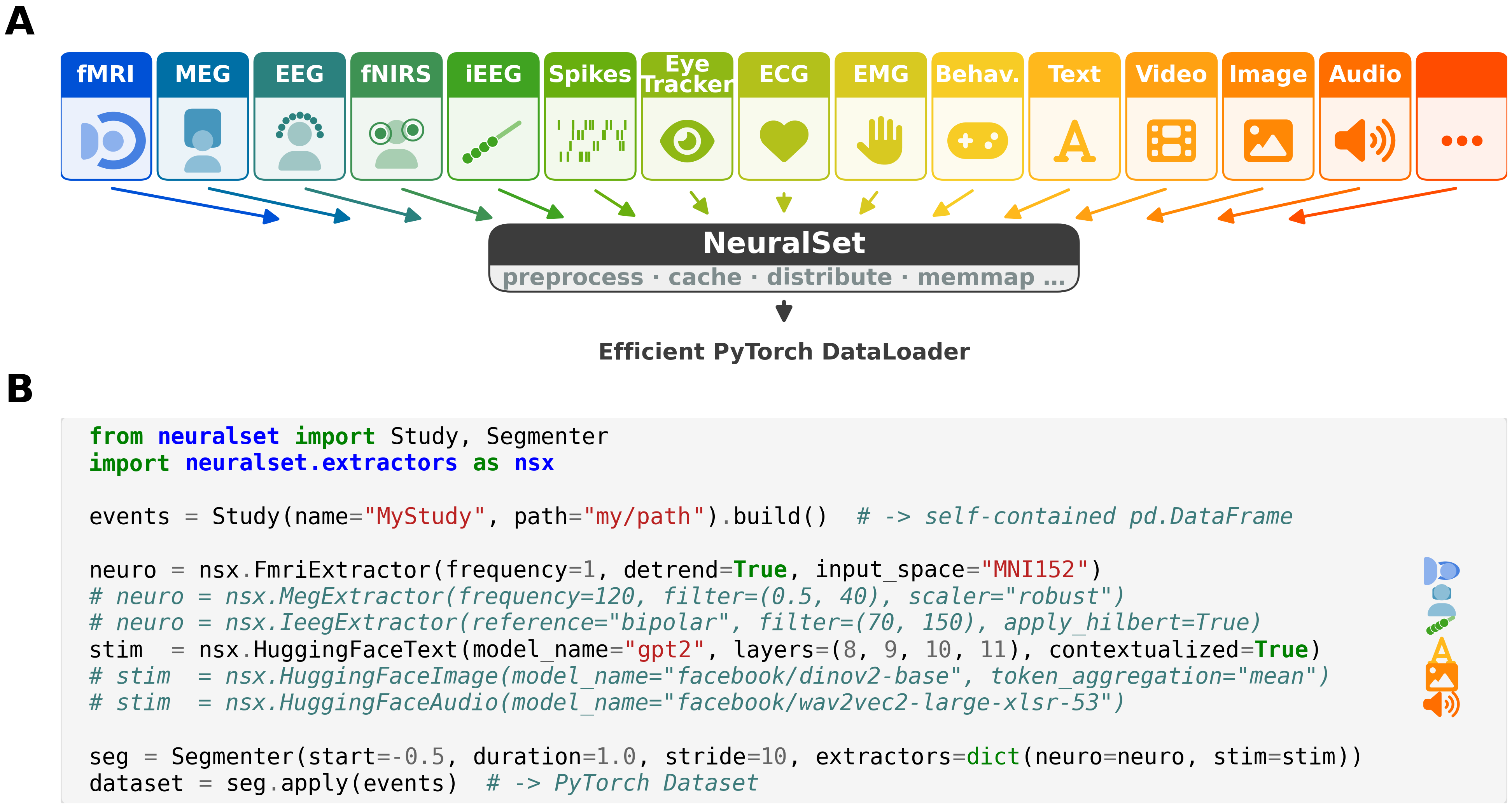}
\caption{\textbf{A.}~NeuralSet unifies the processing of neural recordings and the AI embedding of experimental conditions into a single PyTorch DataLoader.
\textbf{B.}~Specifying the extraction of any types of data requires changing the configuration of a single ``extractor''.}
\label{fig:overview}
\end{figure}

\section{Introduction}

\paragraph{The rise of naturalistic Neuro-AI.}
Neuroscience is undergoing a paradigm shift, transitioning from highly controlled, small-scale experiments 
to massive, multi-modal, and naturalistic datasets \citep{hasson2004intersubject, sonkusare2019naturalistic}. 
This evolution is essential for understanding the biological foundations of intelligence in real-world contexts 
and effectively fuels the emergence of new discipline: Neuro-AI \citep{richards2019deep, yamins2016using}. 
Yet, the application of deep learning onto neuroscientific data 
\citep{caucheteux2022brains, defossez2023decoding, schrimpf2020artificial} is severely bottlenecked 
by a fragmented software ecosystem, which leads most labs across the world, to re-implement similar home-made data processing workflows. 
As public datasets reach the terabyte scale\footnote{\url{https://openneuro.org/}}  
and experimental protocols increasingly incorporate complex stimuli 
like continuous speech and video \citep{allen2022massive, gwilliams2023introducing, nastase2021narratives}, the infrastructure required to process these data has failed to keep pace.

\paragraph{A rich but fragmented software landscape.}
Over the past two decades, the neuroscience community has built a large set of open-source tools 
that have standardized analysis practices and enabled thousands of studies. 
Established softwares such as MNE-Python \citep{gramfort2013meg}, 
EEGLAB \citep{delorme2004eeglab}, FieldTrip \citep{oostenveld2011fieldtrip}, 
and Brainstorm \citep{tadel2011brainstorm} for electrophysiology, 
or Nilearn \citep{abraham2014machine} and fMRIPrep \citep{esteban2019fmriprep} for neuroimaging, 
are the gold standard for signal processing and statistical inference. 
However, these tools were originally designed for a pre-deep-learning era: 
they typically rely on eager loading -- assuming datasets can fit entirely into RAM -- and lack native 
abstractions to temporally align neural time series with high-dimensional, pretrained embeddings 
from modern AI frameworks (e.g., HuggingFace transformers \citep{wolf2020transformers}). 
Consequently, researchers must build ad-hoc pipelines that require manual data wrangling, 
manual caching, and complex backend configurations.

\paragraph{Structure--data decoupling.}
To bridge this gap, we introduce NeuralSet, 
a Python package designed to efficiently unify the processing of diverse neural recordings 
and their associated experimental conditions. 
NeuralSet is built upon the principle of structure-data decoupling: 
it represents the logical structure of any experiment as lightweight, event-driven metadata, 
separating it entirely from the memory- and compute-intensive extraction of the underlying signals. 
This architecture enables lazy, on-demand data processing, allowing researchers to filter, recombine, 
and iterate on massive datasets without loading a single byte of raw data into memory 
until it is required by a PyTorch dataloader \citep{paszke2019pytorch}.

\paragraph{A unified, scalable interface.}
By providing a single, backend-agnostic interface, NeuralSet harmonizes the validated preprocessing stacks 
of domain-specific libraries with the embedding capabilities of modern AI models. 
It natively supports a comprehensive array of neural modalities: e.g. fMRI, EEG, MEG, iEEG, fNIRS, EMG, 
and spike trains, alongside the embedding of naturalistic text, audio, and video stimuli. 
Critically, its caching and hardware-agnostic execution relies on exca \citep{exca}, which allows researchers to scale seamlessly 
from local prototyping on a laptop to distributed execution on high-performance computing clusters. 
Ultimately, NeuralSet is designed to eliminate the friction of multi-task and multi-device data orchestration, 
providing the scalable and reproducible infrastructure necessary for the next generation of Neuro-AI research.


\begin{figure}[tbp]
\centering
\includegraphics[width=\textwidth]{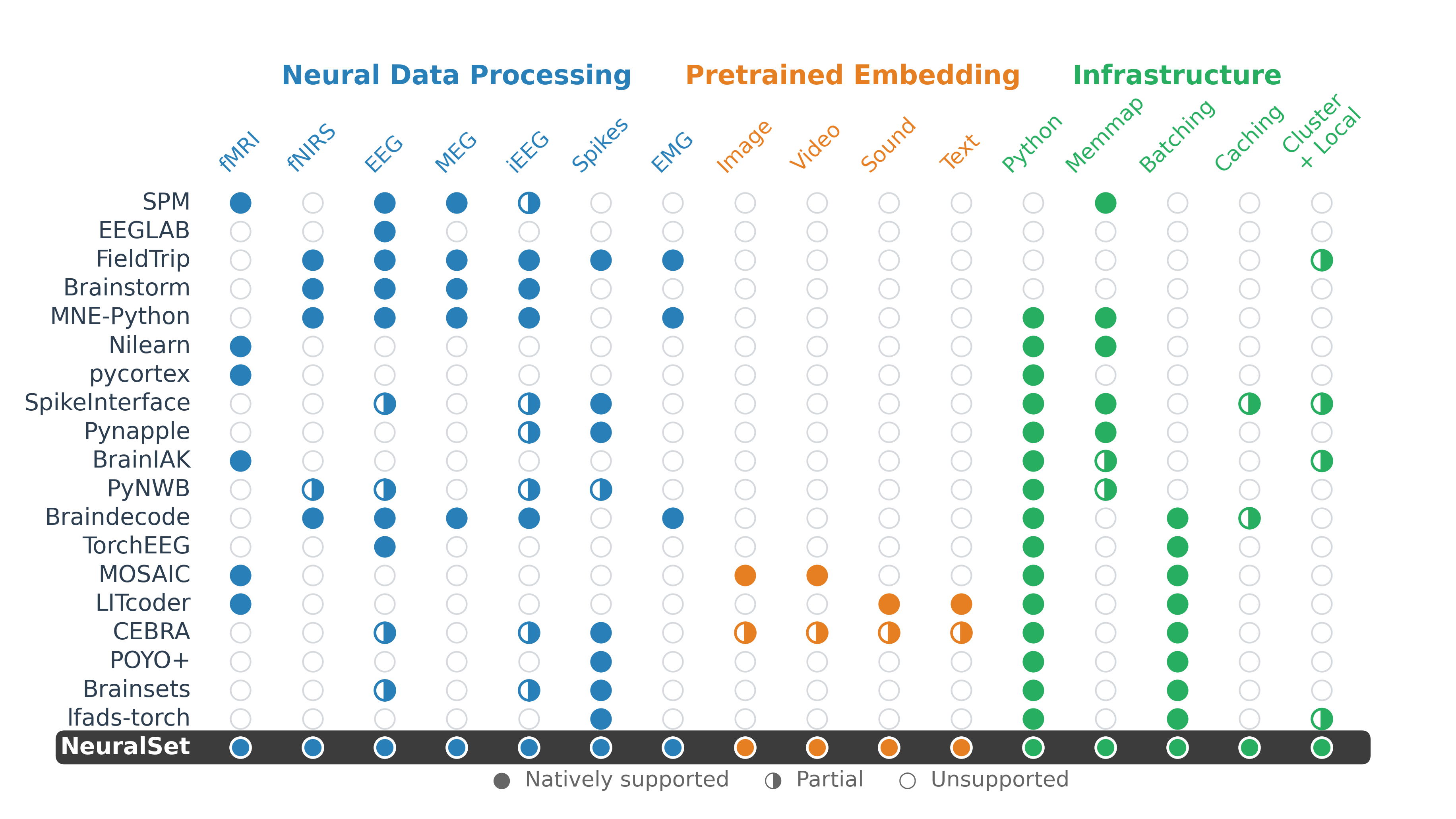}
\caption{Comparison of neuroscience software packages across neural
devices, experimental tasks, and infrastructure features.
Filled circles indicate full support, half-filled circles indicate partial
or compatible support, and empty circles indicate no support.
NeuralSet (bottom row, highlighted) is the only package supporting all
categories.}
\label{fig:comparison}
\end{figure}

\section{Framework}

NeuralSet enforces a separation between the lightweight description of
an experiment and the heavy extraction of data.
The framework is organized around five core abstractions --
\texttt{Events}, \texttt{Extractors}, \texttt{Segments} and \texttt{Tensor Data} -- which
together form a pipeline from raw neural and/or stimulus files to PyTorch-ready datasets.

\begin{figure}[tbp]
\centering
\includegraphics[width=\textwidth]{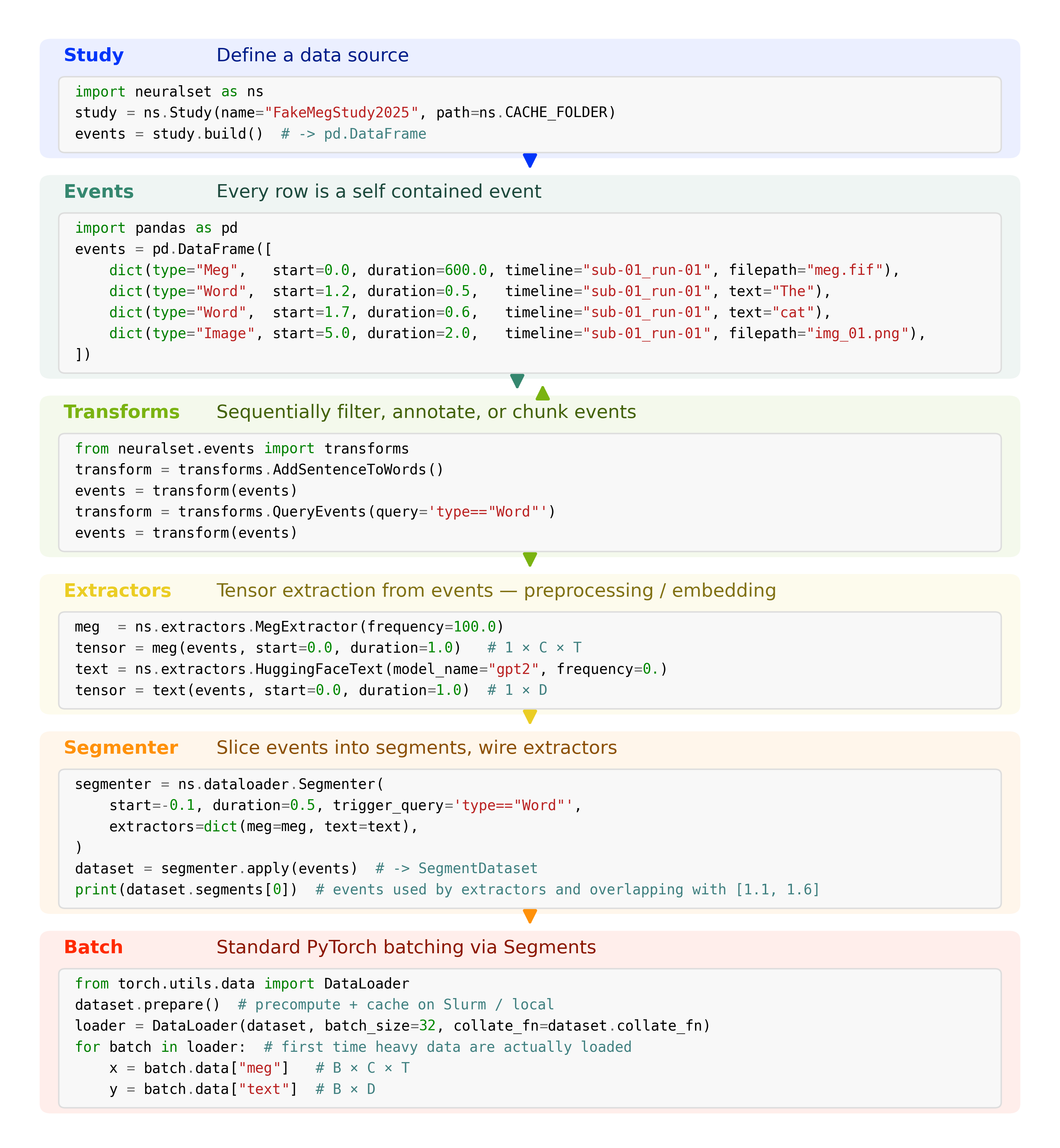}
\caption{The NeuralSet pipeline from user perspective.
Each step is shown alongside its Python code snippet, from defining a
data source (\texttt{Study}) to iterating batched tensors in a standard
PyTorch \texttt{DataLoader}.
The entire pipeline is lazy: only the final iteration triggers data
loading and feature extraction.}
\label{fig:pipeline_code}
\end{figure}

\subsection{Events: describing what happens and when.}

The central insight of NeuralSet is that the logical structure of an experiment
can be represented independently of the underlying data.
Every piece of information -- an fMRI run, a word spoken by a narrator, a
stimulus image -- is modelled as an \emph{event}: a lightweight object (i.e. a Python dictionary)
defined by a \texttt{type}, a \texttt{start} time, a \texttt{duration}, and a
\texttt{timeline} (a unique identifier for a continuous recording session).
For instance, a one-hour fMRI session in which a participant watches a movie is
described by a handful of concurrent events on the same timeline: the fMRI
acquisition, the video stimulus, and potentially hundreds of word-onset
annotations derived from the soundtrack.

A \texttt{Study} object assembles the events of an entire dataset into a single
pandas DataFrame. This approach is consistent with, but not restricted to BIDS-compliant datasets \citep{gorgolewski2016brain}.
Because the DataFrame contains only light metadata -- not the signals themselves --
researchers can explore, filter, and recombine experiments using standard pandas
\citep{mckinney2010data} operations (e.g.\ selecting all timelines of a given subject, or all word
events in a specific language) without loading data into
memory.

Composable \texttt{EventsTransform} operations can then be chained to enrich,
filter, or reorganize events prior to data extraction.
Typical transforms include annotating words with their sentence context,
assigning cross-validation splits, or chunking long audio and video events into
shorter segments.


\subsection{Extractors: turning events into tensors.}

\texttt{Extractors} bridge the gap between the metadata layer and the numerical
arrays required by modern machine-learning models.
Given the events that fall within a temporal window, an extractor reads,
preprocesses, and returns a dense tensor.

For neural recordings, NeuralSet wraps the established preprocessing stacks of
domain-specific libraries.
For example, an fMRI extractor leverages Nilearn \citep{abraham2014machine} for signal cleaning, spatial
smoothing, and surface or atlas-based projection, while an EEG or MEG extractor
delegates to MNE-Python \citep{gramfort2013meg} for filtering, re-referencing, and resampling.
The same interface covers iEEG, fNIRS, EMG, and spike recordings, so that
switching modalities requires only changing a configuration parameter, not
rewriting a pipeline.

For naturalistic stimuli, NeuralSet provides native integration with the
HuggingFace ecosystem.
A single \texttt{HuggingFaceImage} extractor can embed every stimulus frame
through DINOv2 \citep{oquab2023dinov2}, CLIP \citep{radford2021learning}, or any compatible vision model; analogous extractors
exist for audio (Wav2Vec~\citep{baevski2020wav2vec}, Whisper~\citep{radford2023robust}), text (GPT-2~\citep{radford2019language}, LLaMA~\citep{touvron2023llama}), and video (VideoMAE~\citep{tong2022videomae}).
Critically, NeuralSet can expand a static embedding (e.g.\ a single vector per
image) into a time series at an arbitrary frequency, so that stimulus
representations are always temporally aligned with neural recordings.

Extractors follow a three-phase execution model: \texttt{configure} (parameter
validation at construction), \texttt{prepare} (pre-compute and cache heavy
outputs for all events), and \texttt{extract} (lazy retrieval from cache during
model training).
This design ensures that expensive computations --- such as running a large
language model over every word in a corpus --- are performed once and reused
across experiments.

\begin{figure}[tbp]
\centering
\includegraphics[width=\textwidth]{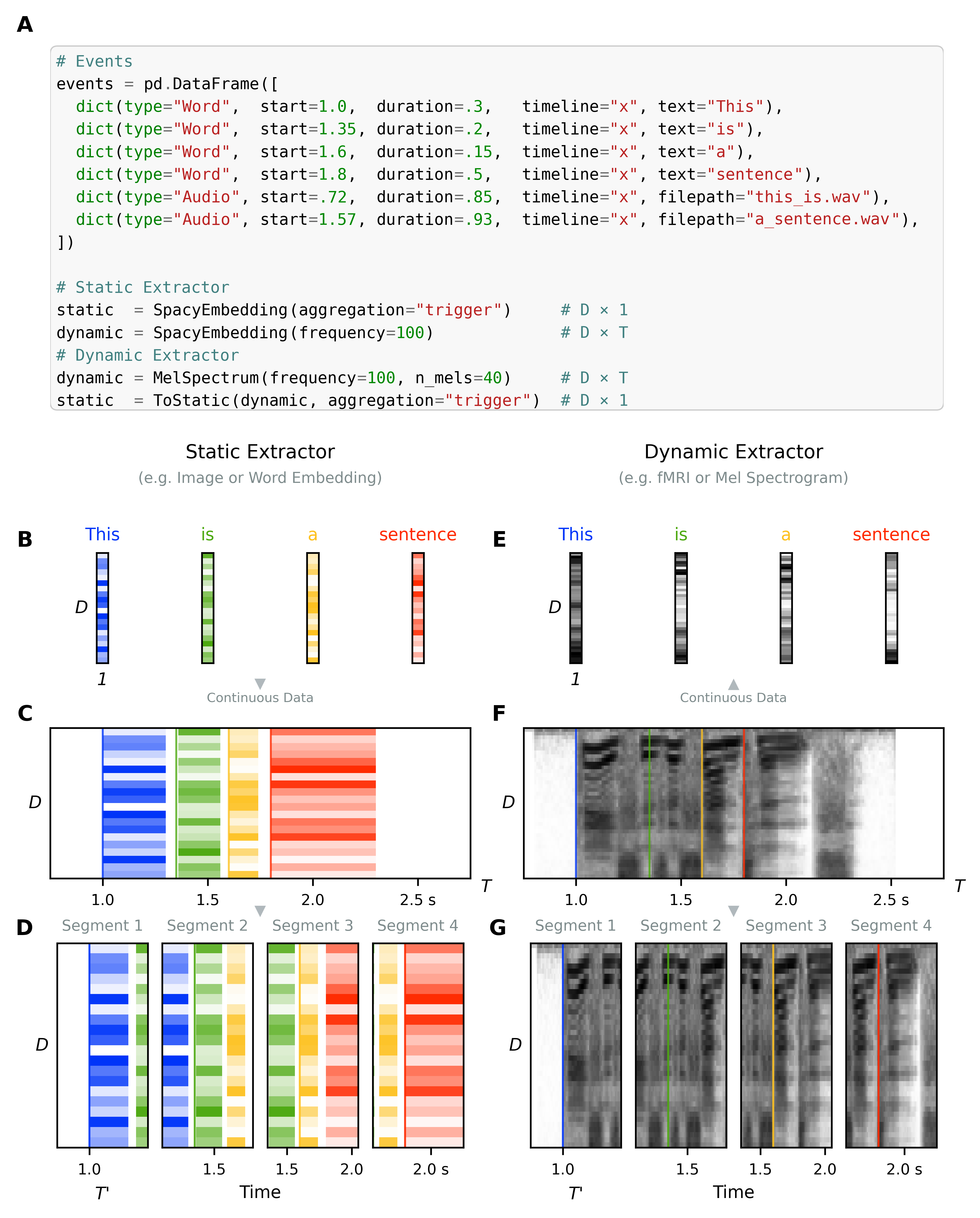}
\caption{Static and dynamic extractors applied to word and audio events.
\textbf{A.}~Python snippet defining the events DataFrame and two
extractors: a static word embedding and a dynamic mel spectrogram.
\textbf{B--D.}~Static extractor: each word is mapped to a single
$D$-dimensional vector~(B), placed on a shared timeline~(C), and
segmented around each trigger onset~(D).
\textbf{E--G.}~Dynamic extractor: each word onset is sliced from the
mel spectrogram into a static $D$-dimensional snapshot~(E), the full
spectrogram is shown on the same timeline~(F), and trigger-aligned
segments are extracted~(G).
Vertical dashed lines mark word onsets; the ``to dynamic'' and
``to static'' arrows illustrate how NeuralSet converts between
representations.}
\label{fig:embeddings}
\end{figure}

A key abstraction is the distinction between the \emph{intrinsic nature} of the
data and the way an extractor \emph{uses} it along the time axis.
Intrinsically \emph{static} data (e.g.\ a word embedding) yields a single
$D$-dimensional vector per event, whereas intrinsically \emph{dynamic} data
(e.g.\ a mel spectrogram) yields a $D \times T$ matrix that varies
continuously over time.
Crucially, NeuralSet allows any extractor to produce its output either
statically or dynamically, regardless of the data's intrinsic nature
(Fig.~\ref{fig:embeddings}).

\subsection{Segments: from events to a PyTorch dataset.}

A \texttt{Segment} is a contiguous temporal window over a set of events,
representing a single training example.
NeuralSet provides a \texttt{Segmenter} that slices the events DataFrame into
segments --- either on a regular grid (sliding window) or anchored to specific
\textit{trigger} events such as image or word onsets.
Each segment retains a reference to its trigger, so that extractors can
distinguish, for example, the particular word that anchored a two-second window
from the other words that happen to fall within it.
Because slicing operates entirely in the metadata domain, millions of segments
can be prepared without touching the raw signal files.

Before any data is loaded, NeuralSet validates that the required events are
present and that all extractors have been prepared, catching configuration
errors early rather than hours into a processing run.
The resulting \texttt{SegmentDataset} is a standard PyTorch \texttt{Dataset}:
it can be wrapped in a \texttt{DataLoader}, passed to PyTorch Lightning, or
consumed by any framework that expects the PyTorch data interface.

\subsection{Batch Data: the actual tensors.}

The output of an \texttt{Extractor} for a single segment, containing a
dictionary of tensors keyed by extractor name, as well as the corresponding segments.
SegmentDataset only contains the list of lightweight segments and the configured extractors.
SegmentDataset.prepare allows to precompute the data for all segments in one go.

\subsection{Backend}

NeuralSet is built upon the \texttt{exca} package -- designed to seamlessly
orchestrate the configuration, validation, and execution of hierarchical
pipelines.

\paragraph{Proactive Configuration Validation.}
To eliminate ``late-stage'' failures -- errors that occur hours into a
processing run, NeuralSet utilizes Pydantic to enforce strict schema
validation at initialization.
For example, if a user specifies a negative filter frequency, or provides an
invalid path for a BIDS directory, the framework raises a validation error
immediately, i.e.\ before job submission.
This proactive check ensures that all parameters, recording modalities, and
hardware constraints are logically consistent before any I/O or
resource-intensive computation begins.

\paragraph{Deterministic and Incremental Caching.}
NeuralSet implements a hierarchical hash-based caching system, where the state
of each pipeline stage is tethered to its specific non-default parameters and
its relevant preceding dependencies.
\begin{itemize}
    \item \textbf{Parameter-Aware Storage:}
    The cache key for a given step (e.g., MEG filtering) is a deterministic
    function of both its local configuration and the upstream data state.
    Changing a single parameter, such as a smoothing kernel width, invalidates
    only the downstream cache, leaving independent branches untouched.

    \item \textbf{Full Provenance:}
    By persisting intermediate results alongside their metadata, NeuralSet
    ensures that any processed tensor can be traced back to the exact version
    of the raw data and the specific preprocessing chain used to generate it,
    facilitating perfect reproducibility across different research
    environments.
\end{itemize}

\paragraph{Hardware Abstraction and Portability.}
A core strength of the framework is its ability to decouple the analysis logic
from the underlying compute infrastructure.
NeuralSet provides a backend-agnostic interface that allows the same script to
be executed across varying environments simply by updating a configuration flag.
When configured for a cluster, the framework automatically handles job
submission, environment synchronization (e.g., Conda/Singularity), and
file path translation.
Because the cache is shared, a researcher can prototype an extractor on a
single local subject and then dispatch the remaining 100 subjects to a
SLURM-based HPC cluster without writing any infrastructure-specific code.

Figure~\ref{fig:comparison} compares NeuralSet with existing packages
across these dimensions.

\tcbset{codepanel/.style={
  colback=white, colframe=black!70,
  fonttitle=\bfseries\scriptsize,
  arc=1.5pt, boxrule=0.4pt,
  left=2pt, right=2pt, top=1pt, bottom=1pt,
}}

\begin{tcolorbox}[codepanel,
  title={\textbf{A.}\ YAML Configuration}]
\begin{Verbatim}[fontsize=\scriptsize]
study:
  name: Gwilliams2022
  path: /data
transforms:
  - name: AddSentences
segmenter:
  trigger_query: "type == 'Word'"
  extractors:
    meg:
      frequency: 120
      filter: [0.5, 30]
    text:
      model_name: gpt2
      frequency: 0.
\end{Verbatim}
\end{tcolorbox}

\smallskip\noindent
\begin{minipage}[t]{0.485\textwidth}
\begin{tcolorbox}[codepanel, equal height group=codepanels,
  title={\textbf{B.}\ Hierarchical Configuration}]
\begin{Verbatim}[fontsize=\scriptsize]
import pydantic, yaml
from neuralset import Study
from neuralset import EventTransform
from neuralset import Segmenter
from neuralset.events.transforms import AddSentences

class DatasetBuilder(pydantic.BaseModel):
    study: Study
    transforms: list[EventTransform]
    segmenter: Segmenter

    def run(self) -> SegmentDataset:
        events = self.study.build()
        for t in self.transforms:
            events = t(events)
        return self.segmenter.apply(events)

# Load from YAML config (A)
cfg = yaml.safe_load(open("config.yaml"))
exp = DatasetBuilder(**cfg)
dataset = exp.run()
\end{Verbatim}
\end{tcolorbox}
\end{minipage}%
\hfill
\begin{minipage}[t]{0.485\textwidth}
\begin{tcolorbox}[codepanel, equal height group=codepanels,
  title={\textbf{C.}\ Caching \& SLURM Dispatch}]
\begin{Verbatim}[fontsize=\scriptsize]
import pydantic, exca

class Preprocessor(pydantic.BaseModel):
    smooth: float = 6.0
    infra: exca.TaskInfra = exca.TaskInfra()

    @infra.apply
    def run(self) -> np.ndarray:
        ...  # expensive computation
        return out

# Caching: first call saves, second loads
p = Preprocessor(smooth=6.0,
    infra=dict(folder="/cache"))
p.run()  # computes and caches
p.run()  # loaded from cache

# SLURM: same code, one config change
p = Preprocessor(smooth=6.0, infra=dict(
    folder="/cache", cluster="slurm"))
p.run()  # dispatched if not computed
\end{Verbatim}
\end{tcolorbox}
\end{minipage}


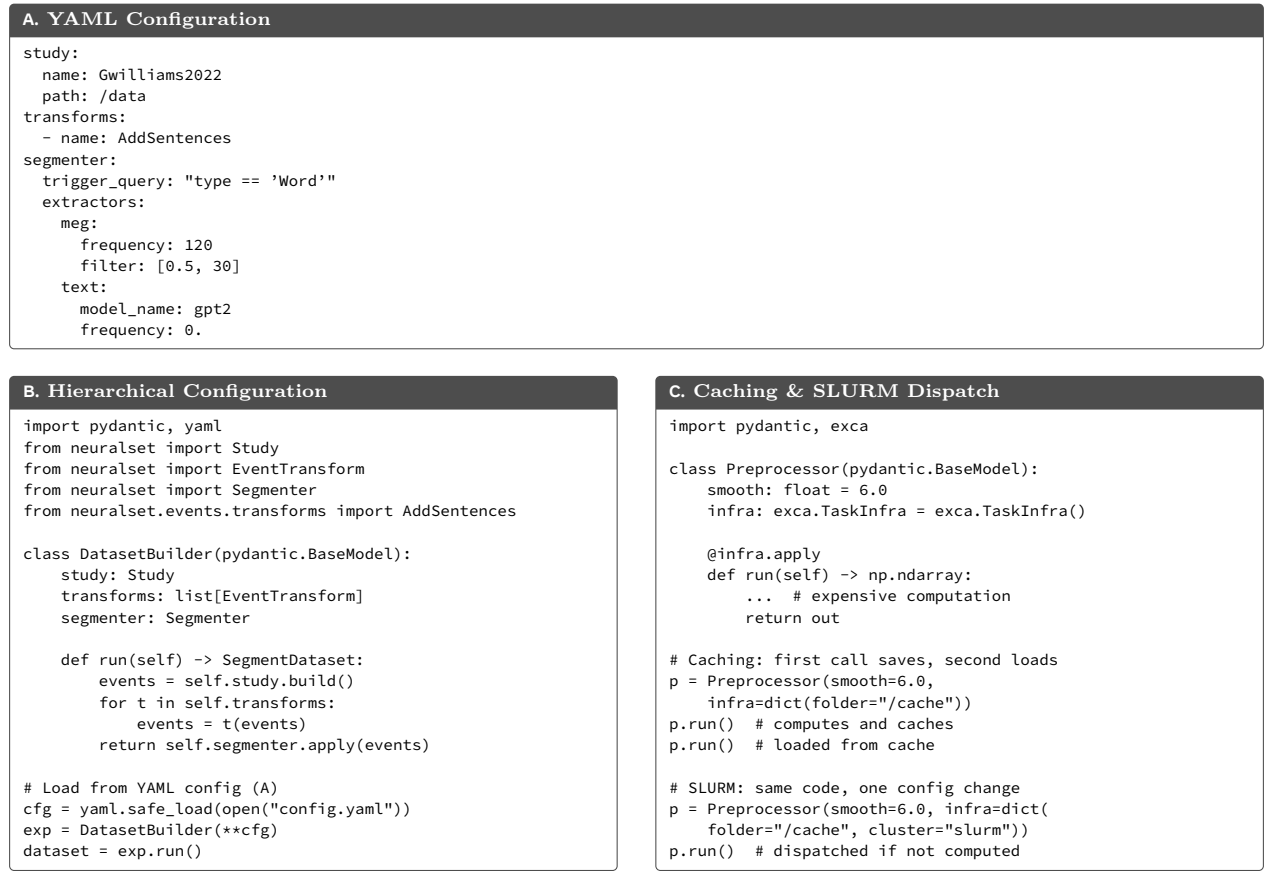
\captionof{figure}{Backend infrastructure.
\textbf{A.}~Example YAML configuration specifying the study, transforms, and
extractors for a complete pipeline.
\textbf{B.}~Hierarchical configuration: every pipeline component is a Pydantic
\texttt{BaseModel}, validated at init and loadable from YAML~(A).
\textbf{C.}~Caching and SLURM dispatch: decorated methods are deterministically
cached on first call; switching \texttt{cluster} to \texttt{"slurm"} dispatches
jobs to HPC with no code change.}
\label{fig:backend}

\section{Discussion}

\paragraph{Summary.}
The quest to understand the biological foundations of intelligence necessitates a fundamental shift 
from fragmented, task-specific analyses toward unified, large-scale modeling of neural dynamics \citep{richards2019deep, yamins2016using}. 
We introduced NeuralSet, a Python framework that bridges the gap between raw, multi-modal neuroscientific recordings 
and modern deep learning architectures. 
By decoupling the logical structure of an experiment from the memory-intensive extraction of its underlying signals, 
NeuralSet provides a scalable, backend-agnostic infrastructure that eliminates the manual data wrangling 
traditionally required in Neuro-AI.

\paragraph{Orchestration, not reinvention.}
Unlike existing software suites that are optimized for modality-specific, eager-loading workflows 
\citep{gramfort2013meg, abraham2014machine}, NeuralSet acts as a specialized orchestrator. 
It does not reinvent validated signal processing algorithms; rather, it harmonizes them with the high-dimensional embeddings 
of contemporary AI models \citep{wolf2020transformers}. 
This intentional specialization ensures that researchers can leverage decades of peer-reviewed preprocessing techniques 
while seamlessly transitioning to PyTorch-ready datasets \citep{paszke2019pytorch}. 
Furthermore, its deterministic caching and hardware-agnostic execution democratize access to high-performance computing, 
allowing complex pipelines to scale from local laptops to SLURM-based clusters without code modification.

\paragraph{Complementing the ecosystem.}
Critically, NeuralSet is not intended to supplant the established tools upon which modern neuroscience rests.
Packages such as MNE-Python \citep{gramfort2013meg}, Nilearn \citep{abraham2014machine}, 
EEGLAB \citep{delorme2004eeglab}, FieldTrip \citep{oostenveld2011fieldtrip}, 
Brainstorm \citep{tadel2011brainstorm}, and fMRIPrep \citep{esteban2019fmriprep} 
embody decades of validated, peer-reviewed signal processing;
NeuralSet neither reimplements nor shadows their algorithms.
Instead, it occupies a complementary niche---the orchestration layer that sits between
these preprocessing stacks and downstream deep-learning frameworks such as PyTorch or Braindecode \citep{schirrmeister2017deep,braindecode}.
Concretely, every signal-processing step is still executed by the original library:
NeuralSet only structures, caches, and batches its output.
The modular \texttt{Extractor} abstraction makes this integration bidirectional:
maintainers of any existing package can expose their pipeline as a NeuralSet backend
through a thin adapter, thereby gaining access to lazy loading, deterministic caching,
and cluster-level execution without modifying their own codebase.
Likewise, the comparison presented in Figure~\ref{fig:comparison} should be read not as a ranking
but as a map of the ecosystem's current coverage:
the gaps reflect the natural specialization of each tool,
and NeuralSet's breadth stems precisely from its ability to delegate to these specialists.
We therefore view this project as community infrastructure and actively invite package maintainers
and domain experts to contribute extractors, propose new modalities, and shape the framework's roadmap.

\paragraph{Limitations.}
Despite its advantages, NeuralSet has limitations. 
Its reliance on underlying libraries (e.g., MNE-Python \citep{gramfort2013meg}, Nilearn \citep{abraham2014machine}) means that it inherits their respective dependencies 
and performance bottlenecks during the initial extraction phase. 
Additionally, while the lazy-loading architecture drastically reduces RAM usage during model training, 
the initial preparation of heavily augmented or high-frequency datasets can still incur significant storage overhead 
due to the caching of intermediate tensors. 
Future development will focus on integrating streaming data formats and expanding native support 
for real-time, closed-loop experimental designs.

\paragraph{Beyond time series.}
In its current form, NeuralSet organizes every recording and stimulus along shared timelines,
making it a natural fit for time-series data such as M/EEG, fMRI runs, and continuous audio or video.
However, many neuroscientific workflows involve interactions between elements
that do not share a common timeline.
For example, MEG source reconstruction typically requires a T1-weighted anatomical MRI
acquired in a separate session, and some structural scans are spatially large volumes
that have no meaningful temporal dimension.
Such cross-session and cross-modality dependencies are not yet first-class citizens in the framework.
We anticipate that NeuralSet can scale to these use cases by leveraging
relational database semantics---for instance through the lazy-frame and join capabilities of
Polars \citep{polars2024}---so that the events dataframe need not be strictly bound
to a single set of timelines but can instead reference auxiliary data through relational links.

\paragraph{Toward cross-modal foundation models.}
Ultimately, the main bottleneck in Neuro-AI is no longer a lack of data---with massive datasets increasingly available 
through standardized formats like BIDS \citep{gorgolewski2016brain}---but the lack of an integrated architecture 
to process it efficiently. 
NeuralSet represents a departure from the era of ``frozen'' datasets and bespoke scripts. 
By unifying disparate domains---from single-unit electrophysiology to whole-brain neuroimaging---under a single, 
extensible framework, we provide the groundwork for neural foundation models that are unconstrained 
by the recording device or the experimental task. 
This infrastructure invites the global computational community to join neuroscientists in the pursuit 
of modeling the biological bases of intelligence.

\section*{Acknowledgements}

We thank Elisa Cascardi, Jennifer Pak, and Pierre Louis Xech for their
steadfast support throughout this project. We are grateful to Alexandre Gramfort,
Arnaud Delorme, Bertrand Thirion, Christophe Pallier, Julie Boyle, Lune Bellec,
Niall Holmes, Pierre Bourdillon, Stanislas Dehaene, Svetlana Pinet, Thomas Moreau,
Valentin Wyart, and Yair Lakretz for their insightful discussions and continued
feedback, and to the broader open-source neuroscience community for the tools
and standards on which this work builds.

\bibliographystyle{assets/plainnat}
\bibliography{main}

\end{document}